# Current-phase Relationship, Thermal and Quantum Phase Slips in Superconducting Nanowires made on Scaffold Created using Adhesive Tape

*Myung-Ho Bae\*, Robert C. Dinsmore III, Thomas Aref, Matthew Brenner and Alexey Bezryadin*

Department of Physics, University of Illinois at Urbana-Champaign, Urbana 61801 Illinois, USA

\*Corresponding Author. E-mail: (M.H.B) mhbae@illinois.edu

**ABSTRACT:** Quantum phase slippage (QPS) in a superconducting nanowire is a new candidate for developing a quantum bit [1,2]. It has also been theoretically predicted that the occurrence of QPS significantly changes the current-phase relationship (CPR) of the wire due to the tunneling between topologically different metastable states [3]. We present studies on the microwave response of the superconducting nanowires to reveal their CPRs. First, we demonstrate a simple nanowire fabrication technique, based on commercially available adhesive tapes, which allows making thin superconducting wire from different metals. We compare the resistance vs. temperature curves of $Mo_{76}Ge_{24}$ and Al nanowires to the classical and quantum models of phase slips. In order to describe the experimentally observed microwave responses of these nanowires, we use the McCumber-Stewart model [4], which is generalized to include either classical or quantum CPR.

Superconducting nanowires (SNWs) have acquired a lot of attention recently due to their potential application in photon detectors [5] and in quantum bits [1,2]. The fundamental interest in SNWs is motivated by the observation of quantum phase slip (QPS) in them [6,7,8,9,10]. The supercurrent in a SNW is determined by the phase difference $\phi$ between two ends of the SNW, and the relationship is known as the current-phase relationship (CPR). Although it has been assumed for many decades that CPR of long wires is multivalued [11,12], recent analysis of this problem that has taken QPS into account has shown that CPR must be single-valued [3,13]. A useful way to probe the CPR is to study the superconducting transport under microwave (MW) radiation [14], which changes the phase difference periodically and produces lock-in resonances, which occur as voltage plateaus on the voltage-current ($V$($I$)) curves, known as Shapiro steps (ShSs) [15,16].

In this Letter, we investigate such resonant behaviors in two different wires: $Mo_{76}Ge_{24}$ and Al nanowires. To fabricate these nanowires, we have developed a simple method based on adhesive tape scaffolds. We measure $V$ ($I$) curves under MWs and observe voltage plateaus on them, which correspond to the phase lock-in resonances. Both of $Mo_{76}Ge_{24}$ and Al SNWs show half-integer resonances, at high enough frequencies (above ~8GHz). We model this phenomenon using the McCumber-Stewart approach [4], by assuming a nonsinusoidal CPR. The best agreement between the models and the data is obtained if we assume that transport properties of $Mo_{76}Ge_{24}$ wires are dominated by thermally activated phase slips (TAPS) and long enough to have a multivalued CPR. Yet, at the same time, $R$ ($T$) curves of $Mo_{76}Ge_{24}$ wires can be fit with theoretical curves originating from QPS theories [17,18]. Thus the possibility of QPS effects influencing $R$ ($T$) curves cannot be excluded. The Al wire sample also shows fractional Shapiro steps. This fact excludes the possibility that the CPR of the wire is non-quantum short-wire limit. We explain the results using quantum CPR due to Khlebnikov [3,13]. Yet the possibility that the wire is not short enough and thus its CPR is close to the multi-valued (i.e. in the intermediate regime) case cannot be excluded definitely.

We observe that when a piece of an adhesive tape is attached to a flat substrate with a trench [19] and subsequently peeled off from the substrate as shown in Figure 1a, some polymer nanostrings are suspended over the trench if the width of the trench is sufficiently small, i.e., a few micrometers wide [see also Supporting Information]. Figure 1b shows a scanning electron microscope (SEM) image of two parallel representative polymer nanostrings suspended over a trench after the "tape-peel-off" process, where the width of the trench is ~1.2 µm. The widths of both strings are ~45 nm. We noticed that narrower trenches gave smaller widths. For example, 20 nm-wide strings have been obtained over an 80-nm-wide trench (Figure 1e).

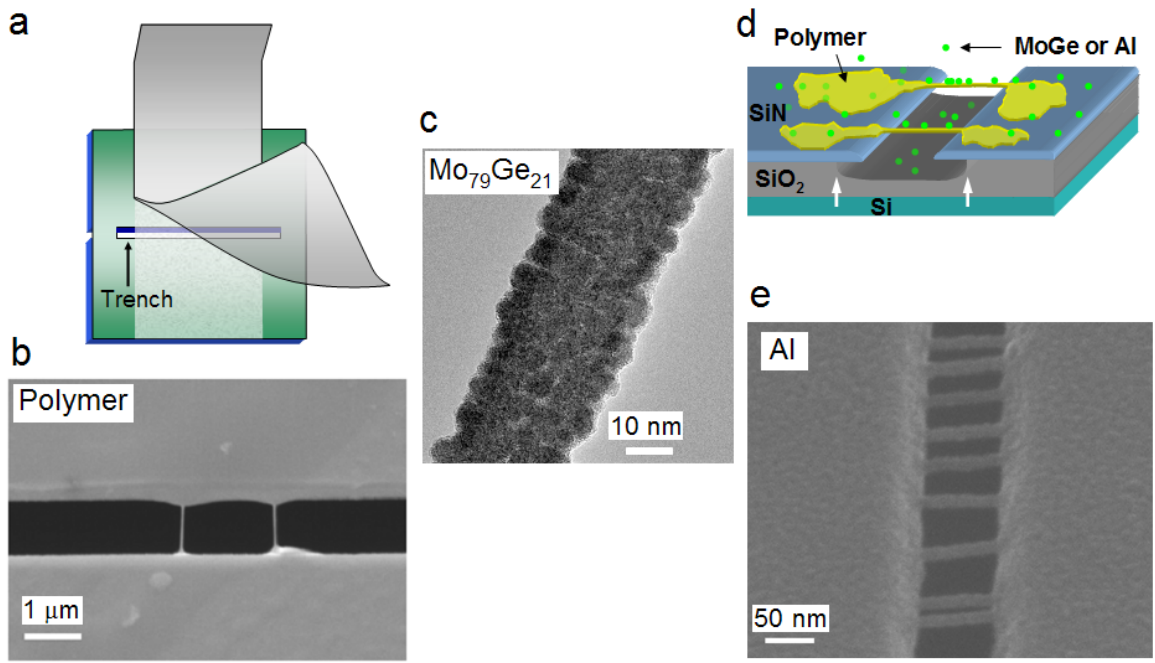

*Figure 1.* (a) Schematic of a "tape-peel-off" technique with an adhesive tape on a substrate having a trench, which is prepared as in Ref. [19]. After the peel-off process, some polymer strings are suspended over the trench. (b) Scanning electron microscope (SEM) image of two polymer strings suspended across a trench, where the width and the length of the strings are 45 nm and 0.8 µm, respectively. (c) High-resolution TEM image of a polymer string after a deposition of a 12 nm thick $Mo_{76}Ge_{24}$ film. The oxidization layer on the surface of the wire shows a different contrast compared to the inside core of the wire. (d) Schematic of a substrate with an undercut-trench with polymer strings,

*where arrows indicate undercuts forming in the SiO$_2$ layer. Metals such as Mo$_{76}$Ge$_{24}$ and Al (green scattered points) are sputtered on the substrate after the formations of nanostrings over a trench. (e) SEM image with a 52° tilt angle, with the axis of the tilt being perpendicular to the trench and lying in the plane of the sample. 25 nm-thick Al nanowires appear over the trench, produced by a deposition of Al. The width of the trench is 80 nm.*

We sputtered amorphous Mo$_{76}$Ge$_{24}$ alloy to the substrate having nanostrings under a pressure of 1×10$^{-7}$ Torr at room temperature. Samples for a high-resolution transmission electron microscope (TEM) imaging were prepared by depositing polymer nanostrings on TEM compatible slits as shown in Figure 1a. Figure 1c shows a high-resolution TEM image of part of the 12 nm-thick Mo$_{76}$Ge$_{24}$ nanowire on a 0.8 μm-long nanostring template, where the width (*w*) of the nanostring is 28 nm. Figure S1d shows the entire TEM image of the nanowire of Figure 1c. Figure S1e also shows that the metal connection from a thin film electrode to the nanowire is very smooth. To perform electronic transport measurements, we fabricated an undercut-trench substrate as shown in Figure 1d, where the undercuts forming at the SiO$_2$ layer are indicated by arrows. After the formation of the nanostrings over the undercut-trench, we used a PDMS contact mask to define thin-film electrodes. Figure 1e shows an SEM image of many Al nanowires in parallel, after sputtering 25 nm-thick layer of Al metal, where all nanowires show nearly the same width of 20 nm. We cut all nanowires, except one, by a focused ion beam and got a sample with one nanowire across the electrodes as shown in the lower inset of Figure 2a. In the sample preparation process with polymer nanostrings, we did not use any chemical process because polymers can be easily damaged by chemical treatments.

The open and closed red circles of Figure 2a correspond to *R* (*T*) curves for two different Mo$_{76}$Ge$_{24}$ nanowires [MoGe1 (right): length=150 nm, *w*=20 nm and MoGe2 (left): length=155 nm, *w*=20 nm], respectively. The details of the experimental setup are explained in the "Methods" section. For MoGe1, $T_{C,film}$ of a 14 nm-thick Mo$_{76}$Ge$_{24}$ film was 6.2 K. Below $T_{C,film}$, the *R* (*T*) curve shows a long flat resistive region, and then it shows one more superconducting transition in the nanowire at $T_C$ ~4 K and the resistance goes below our noise level at $T$~3 K.

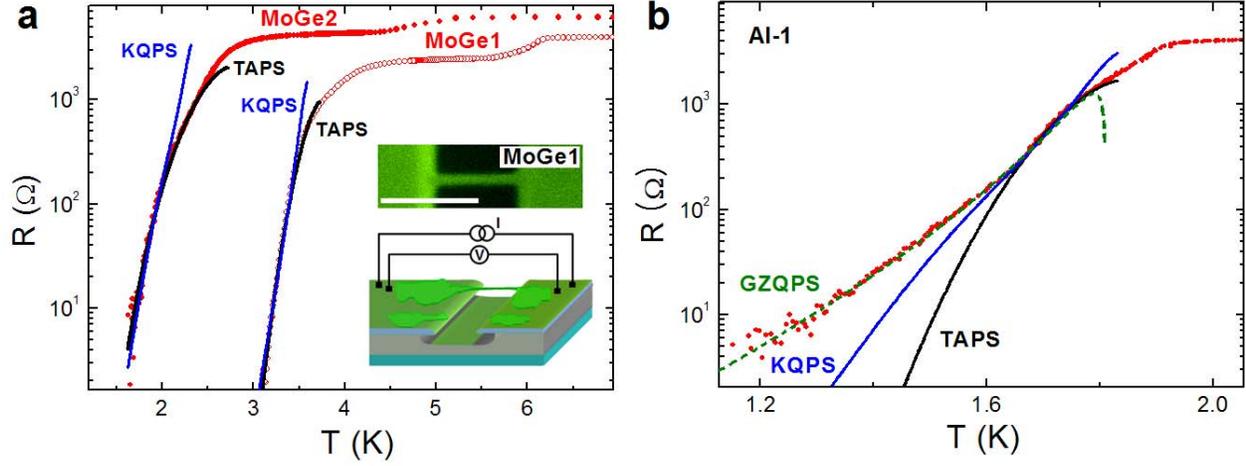

*Figure 2.* (a) R (T) curves of two $Mo_{76}Ge_{24}$ nanowires (MoGe1, open circles and MoGe2, closed circles). The solid black line on each R (T) curve is the best fit produced by Arrhenius-Little (AL) formula $R_{AL}(T)$ (Eq.1), which is due to TAPS, with the three fitting parameters listed in the text. The solid blue lines are obtained from the Khlebnikov QPS theory (KQPS), $R_{KQPS}(T)$ [3,13]. Upper inset: SEM image of wire MoGe1. The scale bar is 200 nm. Lower inset: schematic for the transport measurement setup. The lower nanowire, drawn in Figure 1d, is removed. Focused ion beam is used to remove unwanted nanowires. (b) An R (T) curve of an Al nanowire (Al-1, scattered points). The dashed green curve is the prediction of the Golubev-Zaikin QPS theory ($R_{GZQPS}(T)$) [27] with five fitting parameters (listed in the text) and the solid black curve corresponds to the phenomenological AL fit ($R_{AL}(T)$) [23]. The solid blue line is the best fit predicted by KQPS theory.

In a narrow superconducting channel, a resistance below $T_C$ can be due to TAPS [20,21,22]. A simple approximate semi-phenomenological formula to fit the R (T) curves is the Arrhenius-Little (AL) formula [23,24]:

$$R_{AL}(T) = R_N \exp\left(-\frac{\Delta F(T)}{k_B T}\right). \qquad (1)$$

Here $\Delta F(T)[\approx 0.83[\sigma/\xi(0)](R_Q/\rho)k_B T_c(1-T/T_c)^{3/2}]$ is the energy barrier for phase slips [21], $R_Q$ [=h/4e$^2$=6.45 k$\Omega$] is the quantum resistance, $k_B$ is the Boltzmann constant, $R_N$ [=$\rho L/\sigma$] is the normal resistance of the wire, $\rho$ [~200 µ$\Omega$cm for a MoGe film] is the electric resistivity, $\sigma$ is the cross-section of the wire, and the coherence length $\xi(0)$ is predefined in a dirty limit [25]. The adjustable fitting parameters are $T_C$, $L$, and $\sigma$. The fitting result using Equation (1) is shown by a solid black curve $\xi$ on the R (T) curves of MoGe1 with the fitting parameters: $T_C$ = 3.73 K and $\sigma$ = 250 nm$^2$, and $L$ = 99 nm. The length of the wire measured under SEM was 150 nm. The coherence length was computed from the $T_C$ and was $\xi(0)$=8.8 nm. The fitting is consistent with the experimental results. In the case of MoGe2, the best fit is obtained with parameters of $T_C$= 2.72 K and $\sigma$ = 56 nm$^2$, and $L$ = 57 nm with $\xi(0)$ = 10.3 nm and the actual length of 155 nm. Since the length of Mo$_{76}$Ge$_{24}$ nanowires is sufficiently longer than $\xi(0)$ and $\sqrt{\sigma}$ <4.4$\xi$, these wires are in a quasi one-dimensional long wire limit [11,20,21].

On the other hand, the R (T) curve of the Al nanowire [Al-1, length=80 nm; w=20 nm; the nominal thickness is 25 nm] in Figure 2b shows a long resistance tail, occurring at $T<T_{C,film}$. The AL fit (a black curve) shows a deviation from data below T=1.62 K because the measured tail has a positive curvature. The AL fit always has a negative curvature (in the log-linear plot) because the thermal activation is governed by Arrhenius-type exponential term, like in Figure 2a. In addition, the coherence length 3.5 nm estimated from the TAPS fitting (i.e. using the AL fit) is much shorter than the computed length of (50 ± 20) nm in a dirty limit with $\xi_0$ =1.6 µm and $l$ = (2 ± 1) nm for a ~25 nm thick Al film [26]. Following previous experiments on Al wires [10] we attempt to explain the results using a notion of QPS. In these events the phase slips by 2$\pi$ through a QPS process. The R (T) curve, considering the dissipation by QPS in the Golubev-Zaikin (G-Z) theory [27,10], is

$$R_{GZQPS}(T) = BR_Q S_{QPS}(L/\xi(T))\exp[-S_{QPS}], \qquad (2)$$

where, $S_{QPS}[= A(R_Q/R_N)(L/\xi(T))]$ is the effective action, and $B$ and $A$ are adjustable parameters. The dashed green curve in Figure 2b is the best fit to Equation (2), with $\xi(0)=60$ nm, where the adjustable parameters were $T_C=1.81$ K, $R_N=1.37$ kΩ, $L=80$ nm, $B=5.1$ and $A=3$. Note that in this case we fit the data using only the QPS term, without including TAPS term [10]. The Al-1 fit in Figure 2b is in good agreement with the data, which indicates that the resistance tail could originate from the QPS. Now we compare our results to the experiment of Zgirski et. al. [28]. Their $R(T)$ measurements show that homogeneous wires which are thinner than ~15 nm in diameter can show a pronounced QPS behavior. Homogeneous wires which are thicker than that, say ~17nm in diameter, show only signatures of thermally activated phase slips. Yet non-Arrhenius tails can be observed quite frequently in as-produced Al wires, even in those which have comparatively large diameter of e.g. ~70 nm. Clearly such tails are due to weak links, the presence of which greatly increases the probability of TAPS and possibly QPS. Since our wire is about ~20-25 nm in diameter, most probably the tail we observe is due to some narrow spot in the wire. Yet this narrow segment itself constitutes a weak link which can be analyzed in order to determine whether its resistance is due to TAPS or QPS. On the other hand, since a resistance tail can also originate from external noise or the inhomogeneity of the nanowire [29], one should be careful in the analysis with $R(T)$ curves.

Recently, some theories have predicted that superconducting wires in QPS regime can show a sharp (Arrhenius-type) decrease of the resistance with decreasing temperature [17, 18]. We compare the experimental $R(T)$ curves to such a theory. It was developed by Khlebnikov [18] and it is valid when the QPS rate is low. The resistance as a function of temperature is given by $R_{KQPS}(T) = R_N \exp(-\frac{\pi^2 R_Q \Delta(T)}{4R_N T}\tanh\frac{\Delta(T)}{2T})$ [18], where $\Delta(T)$ is the BCS superconducting gap energy. The fits are shown by sold blue lines in Figure 2. The best fitting parameters were $T_C= 3.6$ K and $R_N=1.7$ kΩ for MoGe1, $T_C= 2.32$ K and $R_N=3.6$ kΩ for MoGe2, and $T_C= 1.83$ K and $R_N=3.2$ kΩ for Al-1. The fits agree well with the MoGe wire data but not with the Al wire data, probably due to a high fugacity of QPS in our Al sample or the presence of weak links. Now we discuss MoGe samples. The measured values for the normal resistance were $R_N$ =2.5 kΩ and 4.3 kΩ for MoGe1 and MoGe2, respectively. Thus we find that $R_N$ used in the KQPS fits is in approximate agreement (same order of magnitude) with the measured $R_N$. These fits, as well as those done by Meiden et al. [17], suggest a possibility that QPS (possibly thermally-assisted QPS), not TAPS, explain the resistance of nanowires at temperatures comparable to $T_C$. But, since the AL fit works also quite well, the possibility of TAPS being the dominant phenomenon cannot be ruled out. Qualitatively different experiments, therefore, are needed to establish whether the TAPS or the QPS is the main contributor to the measured resistance. Below we describe one such experiment, which involves microwave radiation.

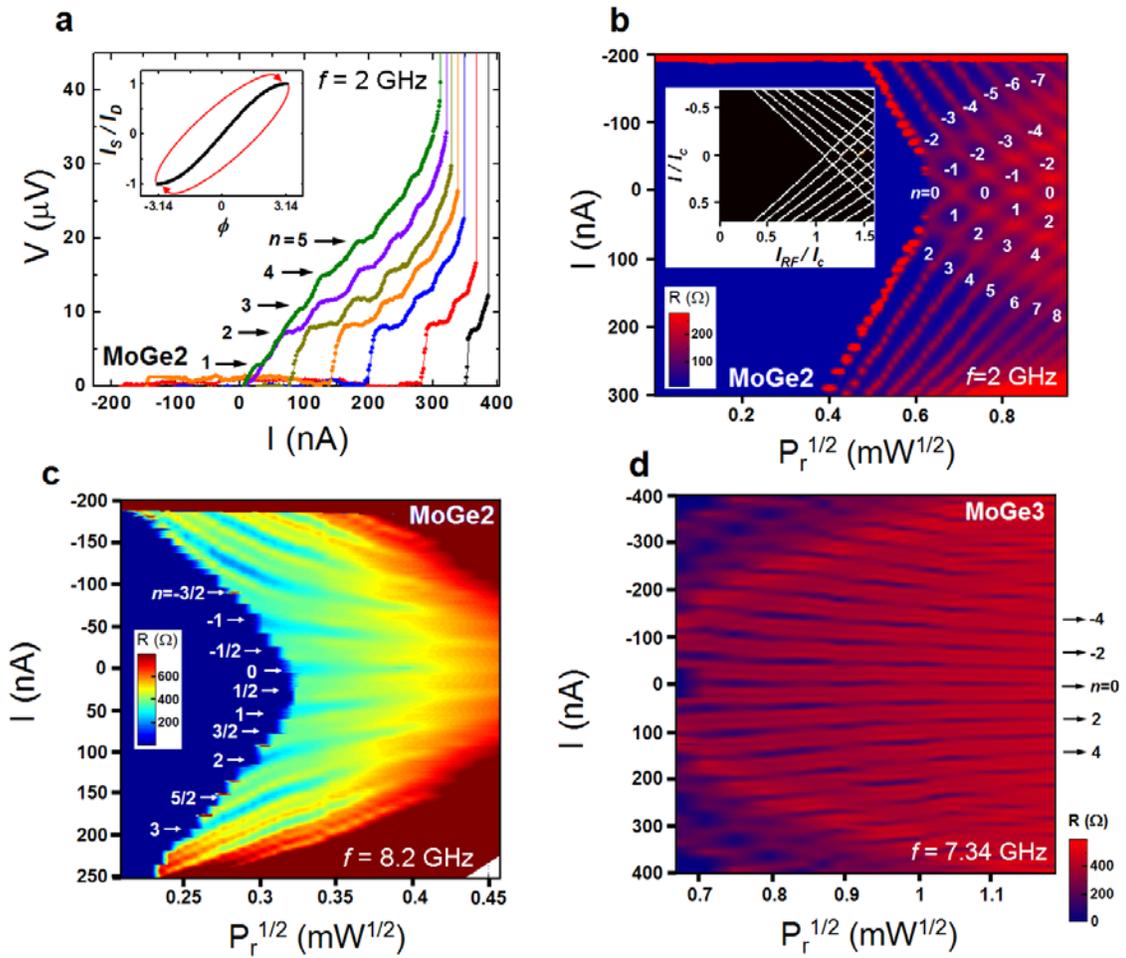

*Figure 3.* (a) V (I) curves of the sample MoGe2 under external MWs of frequency *f=2 GHz* at output power P = -9.5, -8, -6.5, -5.6, -4.8, -4, and -3 dBm, from the right to the left, measured at T=0.32 K. The *arrows indicate the quantized voltage corresponding to multiple integers of the quantum hf/2e=4.1 µV. Inset: An example plot of the LY multivalued CPR [11,12] (note that only one branch is shown) for a nanowire with L/ξ (T)=5.6 (for T=0.32K), where red arrows indicate phase slip phenomena (see text). (b) dV/dI (I, $P_r^{1/2}$) plot for the sample MoGe2, measured at T=0.32 K and f=2 GHz. The large blue-colored region on the left is the supercurrent region. The Shapiro plateaus are also blue, since they are close to being horizontal. They are indicated by index integer numbers. Inset: dV/dI (I/$I_C$, $I_{RF}$/$I_C$) plot, which was calculated with using LY CPR (an example is shown in the inset of (a)), where white colored regions correspond to the differential resistance sharp peaks. (c) dV/dI (I, $P_r^{1/2}$) plot for the sample MoGe2 measured at a higher MW frequency, namely at f=8.2 GHz at T=0.32 K. (d) dV/dI (I, $P_r^{1/2}$) plot of a carbon nanotube-templated nanowire (MoGe3) under MW of f=7.34 GHz at T=0.32 K.*

Figure 3a shows V (I) curves of the sample MoGe2, measured at f=2 GHz with various nonzero powers of the applied MW at T=0.32 K. Each curve exhibits a resistive branch, induced by the MW. We suggest that the observed resistive state is not simply a normal spot on the wire but a coherent dynamic superconducting state, i.e. a microwave-induced phase-slip center (MW-PSC). This suggestion is confirmed by the observation of ShSs (Figure 3a), which represent a phase lock-in effect, which is possible only if the resistive state is phase coherent. The arrows in Figure 3a are positioned at integer multiples of the microwave photon energy, i.e. at $V_n = nhf/2e$ (=n×4.1 µV at f=2 GHz) [30], as expected for ShSs. We find that that the position of arrows matches the position of the voltage plateaus, thus confirming that the resistive state, stabilized by MW in our experiment, is phase coherent. The steps occur as downward peaks in the differential resistance (see figure S2). These peaks represent phase lock-in resonances occurring when the frequency of the revolving superconducting order parameter in the PSC equals an integer multiple of the frequency of the applied MW. Now we consider the current width (ΔI) of the observed steps. Consider for example the second step on the V (I) curve, which appears at V≈8.2 µV. As the MW power is increased, the step first becomes wider but then narrows again and disappears at $P$=-3 dBm ($P$ is defined as the output power of our MW source). Such oscillatory behavior is observed in almost all steps. This trend is even clearer in a $dV/dI$ ($I$, $P_r^{1/2}$) plot of Figure 3b, where $P_r$ is the output MW power expressed in mW units as $P_r = 10^{P[dBm]/10}$ mW. Such plateaus appear as minima (blue color) in the plot and exhibit a pronounced "diamond" structures. Each diamond is marked with an integer number, corresponding to the order of the resonance. The presence of the diamond structure shows explicitly the periodic oscillation of the current widths of the voltage plateaus in the corresponding $V(I)$ curves (Figure 3a).

Now we suggest a model to describe the data of the type shown in Figure 3b. In a long wire limit, Likharev and Yakobson (LY) predicted that the CPR of the nanowire becomes multivalued at $L/\xi(T) \geq 3.6$ [11,12,31]. Such LY CPR has a form of $I_S(\phi) = I_C \frac{\phi \xi(T)}{L_S} \left[ 1 - (\frac{\phi \xi(T)}{L_S})^2 \right]$, which is illustrated in the inset of Figure 3a (a black solid curve) for the MoGe2 wire with $L/\xi(0.3\text{ K}) \sim 5.6$ for $-3.23 \leq \phi \leq 3.23$, which has a multivalued CPR, and $\xi(T) = \xi(0)(1-T/T_C)^{-1/2}$. In the LY CPR exemplified in Figure 3a, the phase slip by $+2\pi$ must occur at $\phi_m$=3.23, for a forward current sweep direction, and at $\phi_m$ =-3.23 (by -2π), for a backward current sweep direction, where the $\phi_m$ points correspond to the maximum supercurrent. The phase slip processes are indicated by red arrows in the inset of Figure 3a. Using this CPR for the phase slip events, we numerically calculated the differential resistance of the wire as a function of a DC current and an AC current ($dV/dI$ ($I/I_C$, $I_{RF}/I_C$)), using the McCumber-Stewart mode [4] at a reduced frequency Ω=0.08 in a resistively shunted case [see method section], where $I_{RF}$ is proportional to $P_r^{1/2}$. The result is shown in the inset of Figure 3b. It is qualitatively consistent with experimental results in the main panel of Figure 3b.

With increasing frequency, $dV/dI$ vs. $2eV/hf$ (differential conductance vs. normalized voltage) curves start to show half-integer steps at $f$=8.2 GHz as shown in Figure S2a [32]. Figure 3c shows $dV/dI$ ($I$, $P_r^{1/2}$) plot measured at $f$=8.2 GHz on MoGe2 at $T$=0.32 K, where half-integer steps with integer steps are shown. At this high frequency we observe a new feature: cyan-colored areas surrounding the yellow regions, which correspond to the voltage plateaus, do not cross and there is no diamonds structure. The oscillation amplitude of $\Delta I$ ($\Delta I$ is the vertical height of a cyan-colored region in Figure 3c, at a fixed power value) with increasing power is also very weak or absent. To check whether these observations are consistent with the multivalued LY CPR in a long wire limit, we numerically calculated $dV/dI$ ($I/I_C$, $I_{RF}/I_C$) plots corresponding to three different situations: a short wire case with $L/\xi(0) \ll 1$ (using a single-valued nonsinusoidal CPR suggested by Kulik and Omelyanchuk [33]) (Figure S3a), and two long wire cases (multivalued nonsinusoidal LY CPRs) of $L/\xi(0) = 5.6$ (Figure S3b), and $L/\xi(0) = 5.9$ (Figure S3c) at $\Omega = 0.8$ in the supporting information. The frequency $\Omega$ was tuned to match the apparent highest order of fractional resonances in the data (Figure 3c and 3d). In the calculation, the current widths corresponding to each voltage plateau in the three cases considered above oscillate with power but do not go to zero at any power. Therefore, this non-zero behavior in $\Delta I$ is a common property of these nonsinusoidal CPRs (we checked explicitly that $\Delta I$ does go to zero at certain values of the MW power if the CPR is sinusoidal). In the simulation, $\delta I_{min}$, the minimum of the current width corresponding to the interval between two resistance peaks for the $n=0$ Shapiro steps, gets larger with increasing $L/\xi$ (see Figure S3c). Thus, the apparent non-zero $\delta I_{min}$ with increasing power by $P_r^{1/2}=0.4$ mW$^{1/2}$ in Figure 3c could be related to this multivalued CPR. However, in Figure 3c the oscillation behavior, predicted by computations (Figure S3c), is not visible. Our explanation is that the examined power is too strong to show a pronounced oscillation behavior, i.e., this system goes to a non-coherent resistive state without resonance steps just above $P_r^{1/2}\sim 0.45$ mW$^{1/2}$. To verify that $\Delta I$ ($P$) can show oscillations with increasing power (as our numerical simulation predicts) we prepared another long wire (MoGe3) using a different method, namely the fluorinated-carbon-nanotube-templating method [19]. The MoGe3 sample was also made of $Mo_{76}Ge_{24}$ alloy. The length of the wire was ~100 nm. Figure 3d shows MWs response ($f$=7.34 GHz) of the MoGe3. The plot is reach in details. The blue color represents the minima of the differential resistance, i.e. corresponds to various orders of lock-in resonances. In particular, half-integer steps are visible as thin blue lines between the thick blue lines corresponding to integer resonances. Most importantly, the width of the integer ShS remains larger than zero for all measured powers, manifested by the fact that the corresponding blue lines do not exhibit interruptions. The $\Delta I$ in current steps corresponding to $n=0, 1, 2$ and $3$ shows weak oscillation behavior with increasing power (as is observed in Figure 3d), which is consistent with calculations in Figure S3c computed for a multiple valued CPR in a long wire limit. Thus we obtain evidence that the QPS rate in MoGe wires is low-enough, so that CPR remains multivalued, i.e. the QPS rate is much smaller than the applied MW frequency.

Now we consider MW response on Al wire (Al-1). With increasing frequency, d$V$/d$I$ vs. $2eV/hf$ (differential conductance vs. normalized voltage) curves start to show half-integer steps near $f$=8.4 GHz as shown in Figure S2b. Figure 4a shows $V$ ($I$) curves of Al-1 without (-∞ dBm) and with MW radiation of various powers ( $f$=8.4 GHz, $T$=0.6 K). Under MWs, the $V$ ($I$) curves show voltage plateaus (indicated by arrows in Figure 4a). The voltage spacing between neighboring voltage plateaus is about $\Delta V$=17.3 μV, which is consistent with the expected $\Delta V = hf/2e = 17.37$ μV. In addition to integer ShSs, half-integer ShSs ($n$=1/2, 3/2, 5/2) are also distinguishable in Figure 4a. Figure 4b shows d$V$/d$I$ ($I$, $P_r^{1/2}$) at the same frequency. The dark green regions indicated by integer numbers in the plot correspond to the voltage plateaus in Figure 4a.

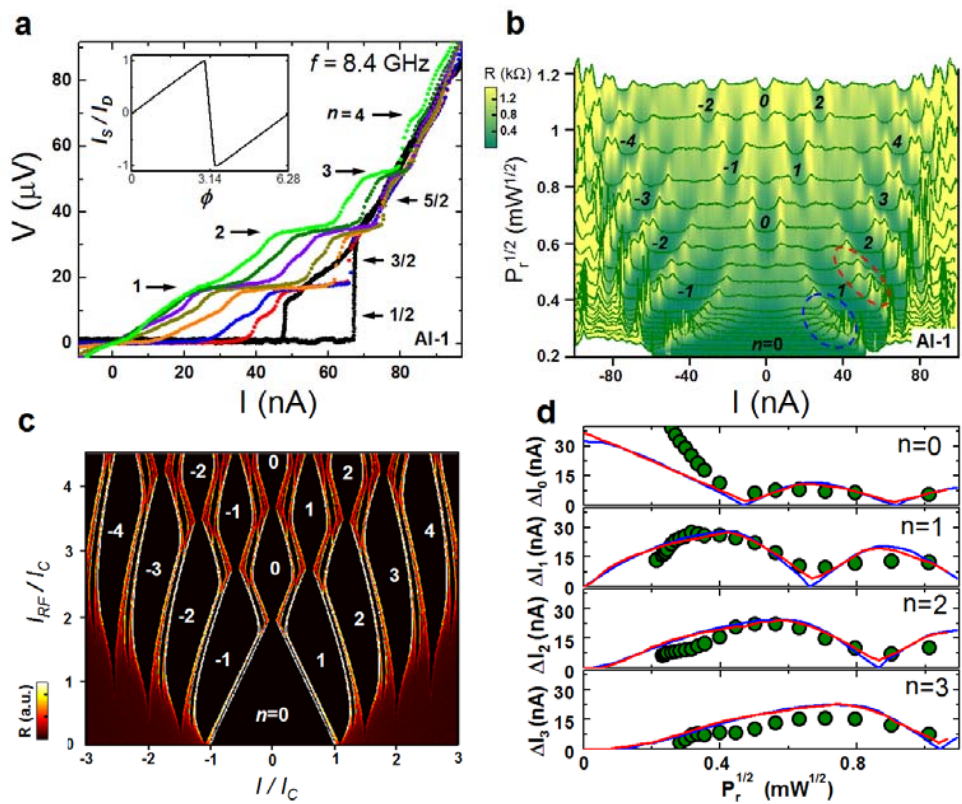

***Figure 4.** (a) V (I) curves of Al-1 under an external MW of f=8.4 GHz at P= -∞, -11.8, -11, -9.5, -8, -7, -6, and -5 dBm from right to left at T=0.6 K. The MW application induces voltage plateaus in V (I) curves. Arrows with integer index numbers (n=1, 2, 3 and 4) indicate voltage positions predicted by the AC Josephson relation, which match with the positions of the observed voltage plateaus. Half-integer Shapiro steps (n=1/2, 3/2, 5/2) are also visible. Inset: An example plot of a QPS-wire theoretical CPR (see text). (b) Differential resistance as a function of a bias current and power, dV/dI (I, $P_r^{1/2}$) plotted for the Al wire, for f=8.4 GHz at T=0.6 K. Solid curves are raw data. The color-coding represents the interpolation of the data shown by the solid curves. The regions shown by dashed blue and red curves indicate examples of half integer ShS (n=1/2 for blue and n=3/2 for red). (c) A dV/dI ($I/I_C$, $I_{RF}/I_C$) plot, numerically calculated by the McCumber-Stewart model, with Khlebnikov quantum theoretical CPR (plotted in the inset of (a), also see supporting information). The integer numbers are index numbers of integer ShS. (d) The closed circles represent current width of voltage plateaus as a function of $P_r^{1/2}$ obtained from (b). The red curves are obtained by numerical calculations of (c) and the blue curves are obtained from Figure S4 (corresponding to classical short-wire CPR, $I_S^{LSW}(\phi)$[11,34]) for n=0,1, 2, and 3.*

In Figure 4b, the local minimum resistance regions between two integer ShSs correspond to half-integer ShSs, e.g., the local minimum resistance regions in blue and red dashed closed loops in the plot correspond to $n$=1/2 and 3/2 half-integer ShSs, respectively. If we consider the temperature dependence of coherence length, we get $\xi$=73 nm at *T*=0.6 K with $\xi(0)$ = 60 nm. This is nearly the same as the wire length of 80 nm. In what follows we will first compare the results with a non-quantum CPR derived for short limit wires and then with a quantum CPR.

In the case of $T_C$~$T_{C,film}$, we can use the Likharev non-quantum (LNQ) short-wire CPR [11,34] as

$$I_S^{LSW} = I_C \left\{ \left[ 1 + \frac{1}{15}\left(\frac{L}{\xi(T)}\right)^2 \right] \sin\phi - \frac{1}{30}\left(\frac{L}{\xi(T)}\right)^2 \sin 2\phi \right\}.$$ An example of this CPR is plotted in the inset

of Figure S4 for $L/\xi$=1.09. It shows only a slight deviation from the sinusoidal CPR. Figure S4 shows a numerically calculated $dV/dI$ ($I/I_C$, $I_{RF}/I_C$) based on the LNQ CPR at $\Omega$=0.6. The current width, $\Delta I$ of the ShS as a function of $P_r^{1/2}$ obtained from Figure S4 (blue curves) agree very well with the experimental result (closed circles) as is shown in Figure 4d [35]. On the other hand, no half-integer ShSs have been found in the calculation (Figure S4). Yet, the experiment shows a pronounced half-integer ShSs in Figs. 4a and 4b. Thus we suggest that LNQ CPR does not apply to our Al wires. Therefore we consider the possibility that QPS is responsible for some observed features. In particular, a resistance tail in Figure 2b suggests existence of QPS in the Al SNW. To incorporate the effect of QPS in the CPR, we use Klebnikov quantum theoretical CPR (KQT CPR) for a QPS-dominated wire, based on the theory of ref. [3,13]: $I_S = \dfrac{I_C \phi}{\pi} - \dfrac{2\alpha^2}{4\pi^3 I_C} \dfrac{\phi}{[1-(\phi/\pi)^2]^2}$ for $-\pi < \phi < \pi$, where $\alpha$ represents the bare QPS fugacity, which is much smaller than unity (also see supporting information). The inset of Figure 4a shows an example of KQT CPR with $\alpha$=0.1$I_C$, which looks like a sawtooth. If $\alpha$ is increased the CPR becomes closer to the sinusoidal one. Figure 4c shows a numerically calculated $dV/dI$ ($I/I_C$, $I_{RF}/I_C$) at $\Omega$=0.6 based on the quantum CPR. The frequency value of $\Omega$ is chosen to make the ratio of the fractional ShS to the integer ShS to be about the same as in the experiment (Figure 4b). Here, the value of $\alpha$ is also chosen to make the ratio $\Delta I$ close to the experimental values. The dark regions indicated by integer numbers in the plot correspond to voltage plateaus of integer ShSs. In the simulation (Figure 4c) half-integer ShSs appear as regions of local minima of the differential resistance (black color), between two neighboring integer steps. The origin of the half-integer ShS is the nonsinusoidal, roughly sawtooth shape of the quantum CPR. It is consistent with the half-integer ShSs observed in the experiments as shown in Figs. 4a and 4b. The consistency is also manifested in their $\Delta I$ as a function of $P_r^{1/2}$ plots in Figure 4d. A general good agreement between the theory and experiment is found except for the case of $n$=0, due to low rates of switching from the static superconducting regime to the dynamic regime. On the other hand, the possibility that the wire is not short enough, i. g., in a case of $\xi(0) \approx 30$ nm with $l \approx 1$ nm and $L = 80$ nm, and thus its CPR is close to the multi-valued (i.e. in the intermediate regime) case cannot be excluded definitely. In addition, since the comparison was done with only one sample for each regime, further experimental evidence is needed.

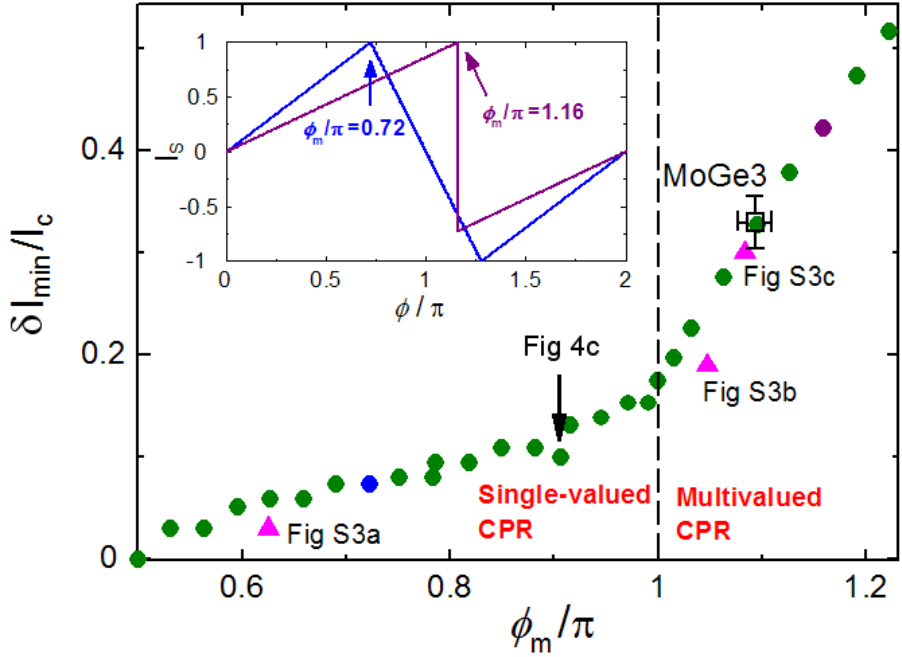

*Figure 5. Scattered solid circles: $\delta I_{min}/I_C$ vs. $\phi_m$, which is obtained by the numerical calculations with various nonsinusoidal CPRs, for example, as shown in the inset, where $\phi_m$ is the phase value corresponding to the critical current in their current vs. phase plots. Solid triangles are obtained by the numerical calculations based on Figure S3. Inset: example plots of the nonsinusoidal CPRs, where plots corresponding to $\phi_m$ =0.72 and 1.16 represent single-valued CPR and multivalued CPR, respectively.*

Since both cases of single-valued (see Figure 4c) and multi-valued CPRs (see Figure S3) show the non-zero $\delta I_{min}$, we plot $\delta I_{min}/I_C$ vs. $\phi_m$ by scattered solid circles in Figure 5 to compare the amplitude of $\delta I_{min}$ with various CPR's in a simplified model version, for example, as shown in the inset of Figure 5. We found that the normalized $\delta I_{min}$ increases faster when $\phi_m$ is larger than $\pi$, at which the system enters the regime of the multivalued CPR (the CPR is single-valued for $\phi_m < \pi$). The solid triangles are obtained from Figure S3 based on exact theories for the CPR (as opposed to our model CPR shown in the insert of Figure 5), which also shows a similar trend with the results based on the simplified model CPR's. In Figure 5, an open square with error bars was obtained from MoGe3 (with $I_C \sim 50$ nA estimated from Figure 2d), corresponding to $\delta I_{min}/I_C \sim 0.33$. The position of this open square corresponds to the multivalued CPR regime. This suggests that the resistance in the MoGe3 could originate from TAPS. For the Klebnikov QPS theoretical CPR plotted in the inset of Figure 4a, $\delta I_{min}/I_C \sim 0.1$ was obtained using by Fig. 4c, which is indicated by an arrow in Figure 5. In the case of Al SNW, since the noise level in $dV/dI$ values was nearly the same a resistance dip between two peaks for $\delta I_{min}$ (not shown), it was difficult to define the value of $\delta I_{min}$. However, we believe that further precise experiments with sufficiently long wires under MWs could reveal the existence of QPS in SNWs, by using the type of analysis presented here.

In summary, we describe a fabrication method to prepare $Mo_{76}Ge_{24}$ and Al nanowires using adhesive tapes on a substrate with ~100 nm wide trench. The widths of the nanowires suspended on the trench are regularly ~20 nm. In the case of the $Mo_{76}Ge_{24}$ nanowires, the phase-slip branch under MWs with a relatively high frequency shows pronounced voltage plateaus corresponding to integer and half-integer ShSs, where power dependence of the width of integer ShSs is explained by the multivalued nonsinusoidal CPR in a TAPS-dominated wire. In particular we observe the step width does not go to zero on any MW power, which indicates that the CPR is multi-valued. Klhebnikov theory of QPS indicated that if QPS is present then the CPR should be single-valued. Thus, we obtain evidence that QPS is not present in the MoGe wires in the studied regime. The Al wire under MWs also shows integer ShSs with half-integer ShSs. We model the observed width using Khlebnikov quantum CPR, which is expected in the QPS-dominated wire, although no direct proof of QPS is found.

## METHODS

### LOW-NOISE MEASUREMENT SETUP

The transport measurements were performed in four-terminal film-involving configurations in a He$^3$ cryostat with the base temperature of 0.28 K. Copper powder-filled epoxy and silver-paste glue at cryogenic temperatures and π-filtering system at a room-temperature are used to suppress the high-frequency noise. All the voltage measurements were done with battery powered pre-amplifiers (PAR113 and SR560). To measure $R$ ($T$) curves we have used a low-frequency bias current with the amplitude of ~ 10 nA and the frequency of 11 Hz, and obtained zero-bias d$V$/d$I$ by fitting a straight line to the linear region of the $V$ ($I$) curve. The microwave signal was introduced through a semi-rigid coaxial cable, with two attenuators at placed at 1K-pot and a copper-helix antenna, which was placed in front of the sample in a Faraday cage and was of the order of 10 mm in size.

### McCUMBER-STEWART MODEL APPLIED FOR NANOWIRES

As the phase slip center (PSC) develops in the wire, some non-zero normal current can flow through the wire. Thus the resistively shunted junction model of McCumber and Stewart can be applied [4]. In this case the phase of the superconducting order parameter can be approximately described by the following phase-evolution equation, $\frac{d\phi}{d\tau} + I_S(\phi) = I_{DC} + I_{RF}\sin(\Omega\tau)$, where $\phi$ is a phase difference between two superconducting electrodes, $\tau$ [$=2\pi f_c t$] is a dimensionless time viable, $\Omega$ [$= f/f_C$] normalized frequency by the characteristic frequency $f_C = 2eI_C R/h$, $R$ the resistance of the resistive branch occurring as a result of PSC, and $I_{DC}$ ($I_{RF}$) are dc bias current (ac bias current induced by an external MW) normalized by $I_C$. Thus the fitting parameters is $\Omega$ and the MW-induced ac-component of the bias current $I_{RF}$ and the dc bias current are running parameters in the numerical simulations. Here, a supercurrent, $I_S(\phi)$ is determined by a CPR of an examined wire, which depends on its physical property as well as the length scale of a wire, as explained in the main text.

**ACKNOWLEDGMENT.** The authors appreciate valuable discussion with S. Khlebnikov, J.-S. Ku and M. Sahu. This work was supported by DOE Grants No. DEFG02-07ER46453. We acknowledge the access to the fabrication facilities at the Frederick Seitz Materials Research Laboratory. This work was also partially supported by the Korea Research Foundation Grants No. KRF-2006-352-C00020.

[35] A scaling factor, $\beta$, for the power with the relation of $I_{RF} = \beta P_r^{1/2}$ and $I_C$ are used as the two adjustable parameters.

# Current-phase Relationship, Thermal and Quantum Phase Slips in Superconducting Nanowires made on Scaffold Created using Adhesive Tape

*Myung-Ho Bae\*, Robert. C. Dinsmore III, Thomas Aref, Matthew Brenner and Alexey Bezryadin*

Department of Physics, University of Illinois at Urbana-Champaign, Urbana 61801 Illinois, USA

\*Corresponding Author. E-mail: (M.H.B) mhbae@illinois.edu

**SUPPORTING INFORMATION**

When one attaches two adhesive tapes into their glue sides and peels off from each other, bunches of glue strands are suspended across two tapes (Figs. S1a and b). The glue strands of the acrylate adhesive tape is generally composed of hard (high $T_g$), soft (low $T_g$), and functional monomers, where $T_g$ is a glass transition temperature [S1]. Fig. S1c shows a representative example of chain composition of an acrylic adhesive [S1]. The internal strength is provided by the hard monomers such as ethyl and methyl acrylate with high $T_g$. The adhesion property originates from the soft monomers with low glass temperatures such as 2-ethylhyxyl, *n*-butyl, and *n*-octyl acrylate. The acrylic acid and acrylic amide play a role of functional monomers for the specific adhesion to desired object.

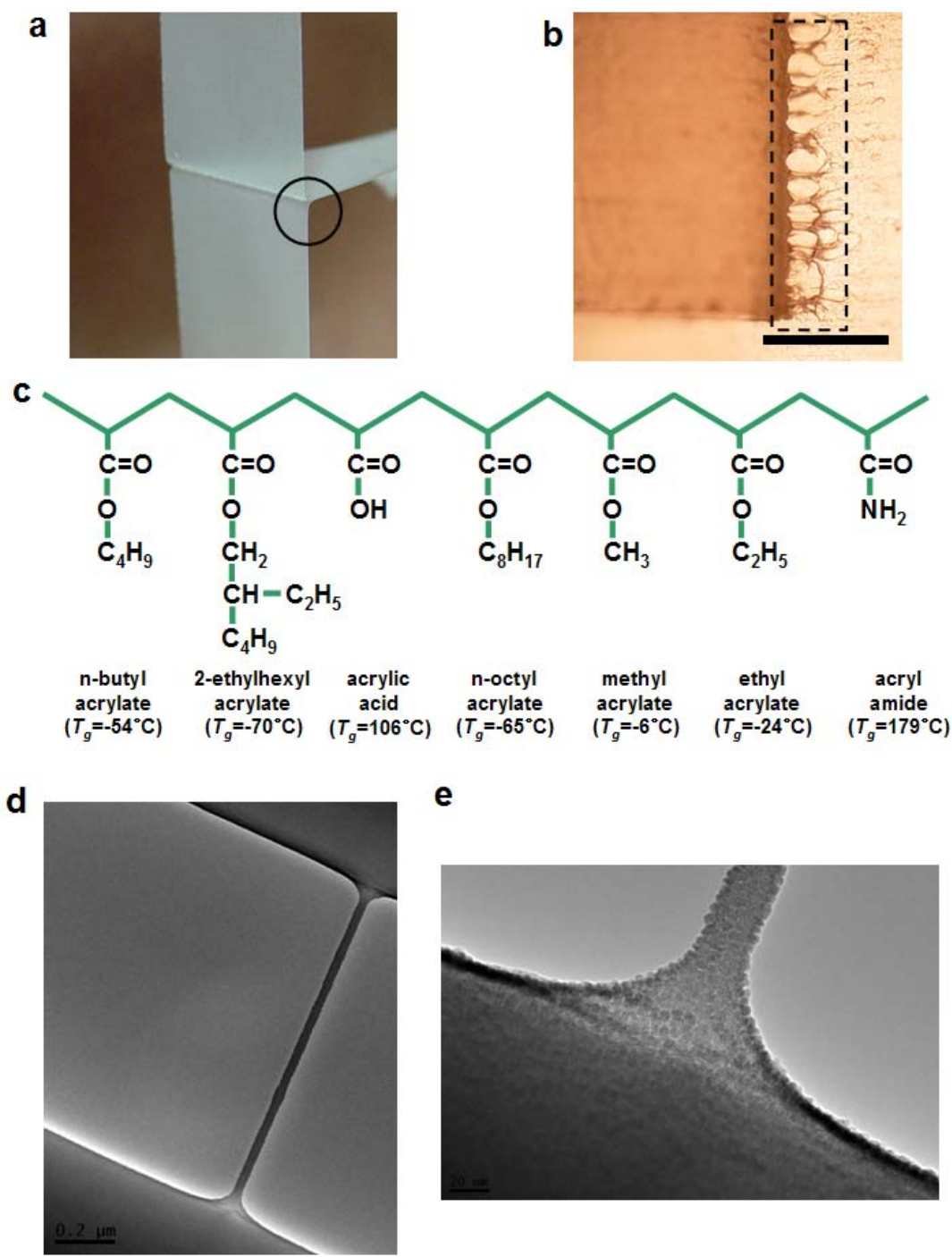

**Figure S1** (a) 3M scotch tape was folded and attached each other. When one tries to separate from each other, one would find polymer strings at an interface such as a black circled area between two tapes. These strings are shown in (b) at the interfaces between two tapes as indicated by a dashed box. Scale bar is 1 mm. (c) Schematic of the representative chain composition of an acrylic adhesive, where $T_g$ is a melting temperature of the monomers from Ref. [S1]. This is not for a 3M Scotch tape. (d) TEM total image of the 12 nm-thick-$Mo_{76}Ge_{24}$ wire in Fig. 1c. (e) TEM image of the junction part between the thin film and the nanowire from another 12 nm-thick-$Mo_{76}Ge_{24}$ wire.

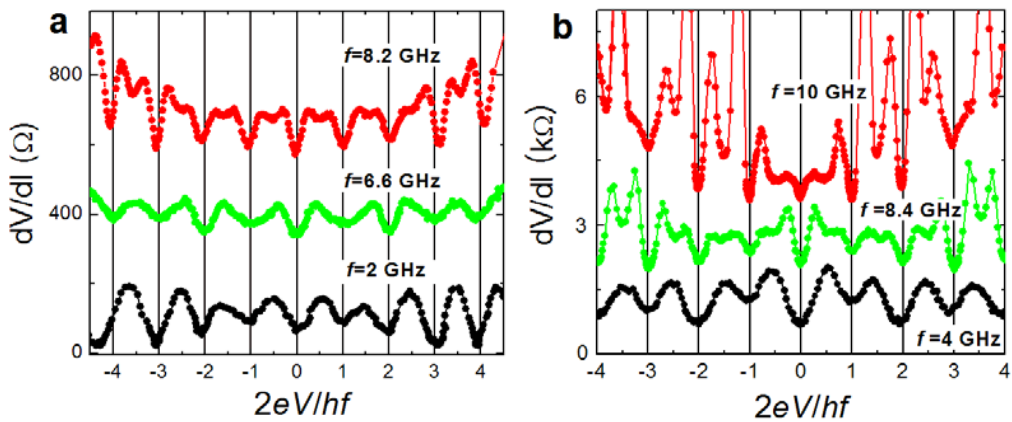

**Figure S2.** (a) $dV/dI$-$2eV/hf$ curves with various frequencies of MoGe2, where 8.2 GHz-one is shifted vertically as 300 Ω for the clarity. (b) $dV/dI$-$2eV/hf$ curves with various frequencies of Al wire, where data are shifted vertically as 1.8 kΩ for the clarity

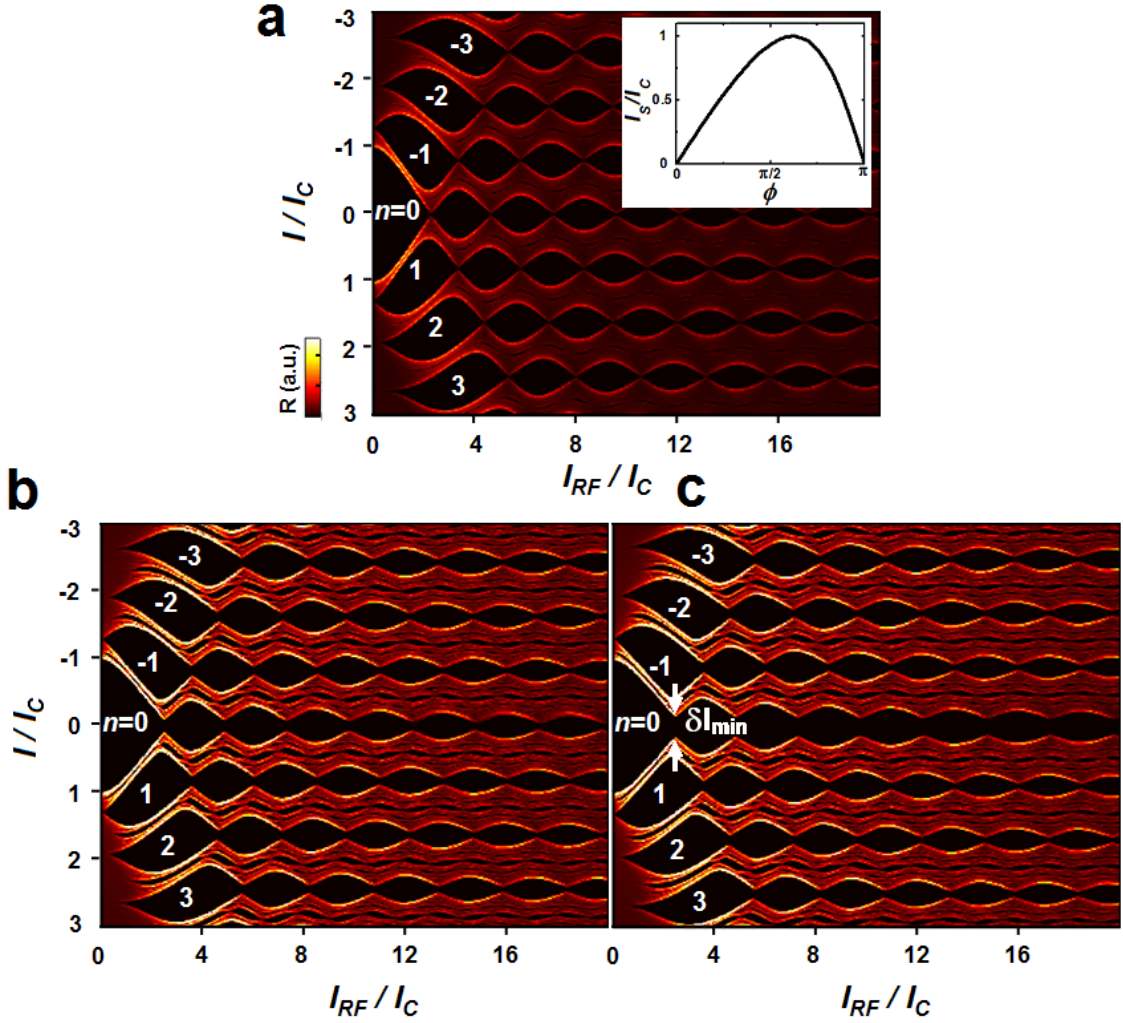

**Figure S3.** d$V$/d$I$ ($I$/$I_C$, $I_{RF}$/$I_C$) plot numerically calculated by the McCumber-Stewart model based on (a) single-valued nonsinusoidal CPR as shown in the inset and multivalued nonsinusoidal CPR with (b) $L/\xi(0)$=5.6 (c) $L/\xi(0)$=5.9 at $\Omega$ =0.8. The integer numbers are index numbers of integer Shapiro steps. Inset of (a): An example of the CPR in the case of a short wire limit. To produce this, we used Kulik and Omelyanchuk model for a diffusive wire [S2]:

$$I_S(\phi) = \frac{4\pi T}{eR_N}\sum_{\omega>0}\frac{\Delta\cos(\phi/2)}{\delta}\arctan\frac{\Delta\sin(\phi/2)}{\delta},$$ where $\delta = \sqrt{\Delta^2\cos^2(\phi/2)+\omega^2}$, $\Delta$ is the

superconducting gap energy, $\omega = \pi k_B T(2m+1)$ are the Matsubara frequency, and $m$ is integer number.

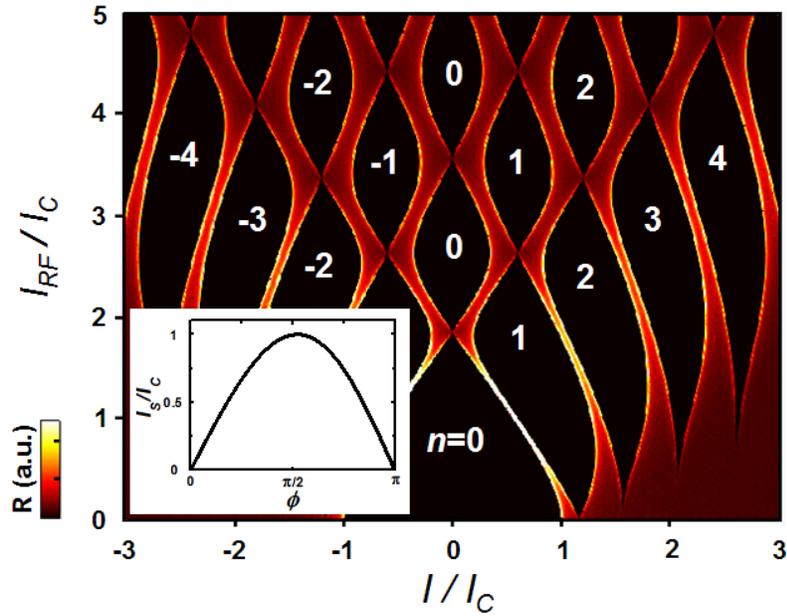

**Figure S4.** $dV/dI$ ($I/I_C$, $I_{RF}/I_C$) plot numerically calculated based on the CPR in the case of $L$[=80 nm]~$\xi(T)$ [=73 nm] as shown in the inset at $\Omega$=0.6 (see main text).

**Current-Phase relation from Mathieu equation in a QPS-dominated superconducting wire**

In this note, the phase difference of the wire is $\phi$, and the charge transported through the wire is $\theta$. The Hamiltonian [S3] is

$$H = -K/2(\partial^2/\partial\theta^2) - \alpha \cos\theta$$

and the energies are given by the Schrödinger equation $H\Psi = E\Psi$.

Make a change of variable $\theta = \pi + 2z$. Then, the Schrödinger equation becomes of the standard Mathieu form:

$$(\partial^2 / \partial z^2)\Psi + (a - 2q\cos 2z)\Psi = 0.$$

Here, $a = 8E/K$ and $q = 4\alpha/K$. Mathieu functions are of the form

$$F_\nu(z) = e^{i\nu z} P(z).$$

Here, $P(z)$ is periodic with period $\pi$. The characteristic exponent $\nu$ is related to the phase difference $\phi$ on the wire by $\nu = \phi/\pi$. So, $\phi = \pi$ corresponds to $\nu = 1$.

A good superconductor corresponds to $q \ll 1$. In this case, the critical current corresponds to $\nu \approx 1$ and is approximately $I_C \approx K/4\pi$ (in units of $2e$). Expansion of $a$ in $q$ is [Abramowitz and Stegun, Eq. 20.3.15]

$$a = \nu^2 + q^2 / 2(\nu^2 - 1) + O(q^4).$$

This is useful as long as $\nu$ is not too close to 1, i.e., the current is not too close to critical current. This gives energy as a function of $\phi$:

$$E = (K/8)(\phi/\pi)^2 + \alpha^2 / \{K[(\phi/\pi)^2 - 1]\} + O(\alpha^4/K^3).$$

Differentiating with respect to $\phi$, we obtain the current in units of $2e$

$$I \approx (K\phi/4\pi^2) - (2\alpha^2/\pi^2 K)\{\phi/[1 - (\phi/\pi)^2]^2\}.$$

## Supporting Information References